# Optimization Strategies for Beam Direction and Dose Distribution Selection in Radiotherapy Planning

Keshav Kumar K.[1*], Dr N V S L Narasimham[2], Dr A Rama Krishna Prasad[3]

[1]Department of Mathematics, Jawaharlal Nehru Technological University, Hyderabad-500 085, Telangana State, India. Email- keshav.gnits@gmail.com
Orcid ID: https://orcid.org/0000-0002-9211-2960

[2]Department of Humanities and Mathematics, G. Narayanamma Institute of Technology and Science (for Women), Hyderabad 500 104, Telangana State, India. Email- nvsl.narasimham@gnits.ac.in
Orcid ID: https://orcid.org/0000-0002-3572-9403

[3]Department of Mathematics, Jawaharlal Nehru Technological University, Hyderabad-500 085, Telangana State, India. Email-prof.prasadark@gmail.com

*Corresponding Author: keshav.gnits@gmail.com

*Abstract*— Radiotherapy planning is a critical aspect of cancer treatment, where the optimal selection of beam directions and dose distributions significantly impacts treatment efficacy and patient outcomes. Traditionally, this process involves time-consuming manual trial-and-error methods, leading to suboptimal treatment plans. To address this challenge, optimization strategies based on advanced artificial intelligence (AI) techniques have been explored. This paper presents an investigation into the application of AI-driven optimization methods for beam direction and dose distribution selection in radiotherapy planning. The study proposes an approach utilizing Convolutional Neural Networks (CNN) to learn the relationship between patient anatomy and optimal beam orientations. The CNN model is trained on a dataset comprising anatomical features and corresponding beam orientations, derived from a column generation (CG) algorithm. Additionally, Particle Swarm Optimization (PSO) and Grey Wolf Optimization (GWO) algorithms are employed to optimize the CNN's weights and biases to attain the Fluence Map Optimization (FMO) objective function. Experiments are conducted using data from 70 clinical prostate cancer patients. The results demonstrate the effectiveness of the CNN-PSO and CNN-GWO approaches in generating beam orientations that yield treatment plans with dose distributions comparable to those obtained through traditional CG. DVH analysis of the resulting plans for different anatomical structures validates the accuracy and feasibility of the CNN-GWO model in radiotherapy planning. The findings of this study highlight the potential of AI-driven optimization strategies to revolutionize radiotherapy planning by significantly reducing planning time and enhancing treatment plan quality.

## I. INTRODUCTION

External Beam Radiation Therapy (EBRT) is a highly prevalent treatment approach employed for diverse cancer types [1]. It harnesses the power of high-energy radiation to target and damage cancerous cells. The primary goal of EBRT is to eradicate cancer cells while preserving the health and functionality of surrounding healthy tissues and critical structures. As the radiation beams pass through the body, they cannot discriminate between cancerous and healthy cells, which can potentially result in harm to normal tissues and organs. The primary objective is to develop a treatment plan that maximizes the impact on cancer cells while simultaneously reducing toxicity and preserving the well-being of healthy tissues [2]. This task poses a significant challenge in Radiation Therapy (RT), as it necessitates finding the right balance between effectively targeting cancerous cells and safeguarding surrounding healthy tissues from unnecessary harm.

Treatment planning is a crucial aspect of EBRT that involves careful consideration and calculation of radiation delivery [3]. Since the radiation beams do not discriminate between healthy and cancerous cells, an effective treatment plan must account for the unique characteristics of each patient's tumor and the surrounding healthy structures. The ultimate goal is to optimize the dose distribution so that cancer cells receive a targeted and effective dose while minimizing exposure to healthy organs and tissues. Despite technological advancements in RT, the integral dose delivered to the patient's body remains constant. The planner's crucial role is to strategically direct the excess radiation to target areas, maximizing its impact on cancer cells while minimizing harm to healthy tissues. This optimization ensures effective treatment with minimal side effects. The focus shifts to achieving the ideal balance between therapeutic benefit and minimizing harm. Each patient's case is unique, and the treatment plan must be personalized to address their specific needs. Advanced imaging techniques, including MRI, CT, and PET scans, play a vital role in precisely delineating the tumor and identifying nearby critical structures. This information helps radiation oncologists and planners develop a treatment plan that is tailored to the patient's anatomy and tumor characteristics. Advancements in RT technology have significantly improved treatment planning and delivery. Sophisticated computer algorithms, including intensity-modulated radiation therapy (IMRT) and volumetric-modulated arc therapy (VMAT), allow for highly conformal dose distributions [4]. These techniques shape the radiation beams to match the tumor's contours while sparing healthy organs as much as possible. IMRT, a sophisticated technique within EBR, has transformed cancer treatment by delivering radiation beams with varying intensities from multiple static directions. This modulation allows precise targeting of the Planning Target Volume (PTV) while sparing nearby healthy tissues. However, the optimization of beam directions, known as the Beam Orientation Optimization (BOO) problem, remains a challenging task. The concept of BOO dates back to 1967 when researchers recognized the impact of beam direction on treatment plan quality. Early attempts involved manual selection of beam angles, which was a time-consuming and sub-optimal process. Over the decades, advancements in technology and computational algorithms have facilitated progress in BOO techniques.

Traditionally, the selection of beam angles was performed manually by radiation planners. This method involved a trial-and-error approach, often leading to suboptimal treatment plans [5]. The increasing complexity of modern treatment plans, with multiple beams and varying intensities, called for automated solutions to streamline the process and improve plan quality. In recent years, researchers have focused on developing automated BOO algorithms that leverage mathematical optimization techniques to find optimal beam configurations efficiently. These algorithms consider various factors, including tumor size, shape, and proximity to critical structures, to determine the best combination of beam directions for each patient. To accurately assess the dosimetric impact of the selected beam orientations, FMO is employed. FMO aims to find the optimal intensity distribution of each beam to create a high-quality treatment plan. Each beam is divided into small beamlets, allowing individual control of their intensities. By combining BOO and FMO, treatment planners can achieve superior dose distributions that effectively target the tumor while minimizing radiation exposure to healthy tissues. Modern BOO approaches attempt to tackle the radiation dosage domain issue by calculating dose influence matrices for all potential beam orientations before addressing the FMO. However, these operations are computationally complex and time-consuming, often requiring hours for dose influence matrices and several minutes for FMO. This hampers BOO's implementation in clinical routines. Efforts to enhance efficiency include developing streamlined algorithms, utilizing parallel computing, and exploring approximation techniques. Improving BOO's efficiency would enable precise and personalized cancer treatments while minimizing the impact on healthy tissues, benefiting patient care in clinical settings.

## II. LITERATURE SURVEY

In the literature survey, various strategies addressing the challenging task of beam direction and dose distribution selection in radiotherapy planning have been extensively explored. These strategies aim to enhance treatment plan quality, reduce treatment delivery time, and improve patient outcomes.

One noteworthy contribution is presented in the journal [6], where a metaheuristic approach combining Genetic Algorithm (GA) and an exact approach is suggested. This approach picks beam positions and calculates dose distributions in IMRT effectively. GA is employed to select a set of beams, and an optimization model is solved using the Interior Point method to evaluate the dose distribution. The results demonstrate that this metaheuristic is not only faster than the exact Branch and Bound methodology but also achieves optimal solutions in multiple experiments, making it a promising approach for practical applications. The study [7] proposes a quick beam orientation selecting technique using a Deep Neural Network (DNN) in the quest for speedier and clinically viable treatments. It is possible to avoid the highly computational calculations of dose influence matrices for every potential candidate beam by training the DNN to identify acceptable beam orientations employing the patient's anatomical attributes. As a result, the proposed method significantly reduces the time required for beam orientation selection, making it well-suited for clinical routines. In the article [8], they see a novel technique, Total-Beam-Space and Beam Angle Optimization (TBS-BAO), for non-coplanar robotic CyberKnife radiation for cancer. Starting with multi-criteria FMO, the approach creates a Pareto-optimal baseline dose distribution. Subsequently, a segmentation and beam angle optimization step is performed to reproduce the ideal dose distribution closely, while restricting the number of allowed beams. This novel strategy efficiently selects optimal beams and showcases the potential for further advancements in non-coplanar radiotherapy techniques. In the domain of noncoplanar IMRT treatment planning, the study [9] proposes the SA-DDL algorithm, incorporating the simulated annealing optimization method along with a pre-trained Neural Network (NN) for dose distribution prediction. The performance evaluation reveals promising results with acceptable dose differences for organs at risk (OAR) and PTV. This reinforces the reliability and potential of deep learning-based optimization for treatment planning tasks.

Additionally, in the article [10], a novel AI-based strategy is presented for managing the combinatorial complexity of finding the ideal ultrasound-robot position and therapy beams. This method considerably improves coverage in ultrasound-guided RT for prostate patients cured with the

CyberKnife over earlier randomized heuristics by selecting candidates for the beams based on CNNs. In the journal [11], a variety of metaheuristic approaches to the beam angle optimization (BAO) problem are analysed and compared. However, the steepest descent algorithm's convergence is too slow for practical use in clinical settings, even though it can find locally optimal beam angle configurations. This research provides a next-descent algorithm that quickly converges to high-quality solutions, providing a realistic and effective method for tackling BAO problems. The author [12] provides a mathematical optimization framework and a metaheuristic termed TSrad based on Tabu Search to tackle the multi-objective nature of beam angle and dosage distribution issues. The TSrad method is designed for large-scale instances and demonstrates the ability to achieve optimality for certain cases and produce viable solutions within reasonable computational time for multi-slice problems, contributing to the advancement of multi-objective optimization techniques in radiotherapy planning. The research [13] introduces a systematic approach for selecting an appropriate number of "more optimal" beam directions without the need for complex objective function optimization or brute force methods. By analysing the relative weights of beamlet contributions, insignificant beams are identified and eliminated from the plan, leading to improvements in target dose uniformity and critical organ sparing. This novel concept exhibits the potential for optimizing beam angle selection while reducing the computational burden associated with exhaustive evaluations. In conclusion, the literature survey highlights a wide range of strategies that hold immense promise for advancing radiotherapy planning practices. These approaches address critical challenges in selecting beam directions and dose distributions, and their integration of AI and optimization techniques promises to revolutionize radiotherapy treatment planning.

### III. METHODOLOGY

AI presents an appealing solution for the BOO problem due to its speed and success in medical applications [14]. CNNs have demonstrated exceptional proficiency in image processing and possess the ability to learn from intricate optimization methods and medical expertise, enabling the provision of personalized treatment plans based on patient-specific anatomical features. This study aims to develop a CNN-based BOO method that is rapid and flexible, yielding results within seconds for seamless integration into clinical routines, thereby expediting cancer treatment planning. The proposed CNN is designed to establish the correlation between patient anatomy and the optimal configuration of beam orientations by taking into account anatomical features and an optimization algorithm. Notably, it can predict beam orientations without the need for prior knowledge of dose influence matrix values. While the research concentrates on a particular objective function and optimization algorithms such as PSO and GWO, the method's adaptability extends to encompass various BOO objective functions and iterative optimization algorithms. To validate its effectiveness, the CNN is trained, assessed, and tested using augmented images of 70 prostate cancer patients, creating a diverse and clinically relevant dataset. The sample image of a CT scan of prostate cancer is shown in Figure 1. The optimization algorithm, CG, selects beams iteratively to improve the solution, followed by FMO for dose distribution. The CNN replaces the CG phase, mimicking its reasoning and internalizing the FMO solution without directly computing beam influence matrices. The feasibility of the network was evaluated using a fixed count of beams (five) for the treatment of patients. In summary, the study introduces an AI-based BOO method, employing CNN to rapidly generate personalized treatment plans without the need for extensive computations of dose influence matrices. This approach streamlines the treatment planning process and has potential applications in various BOO settings.

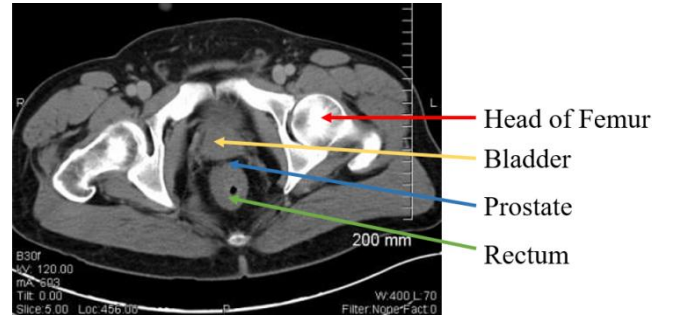

Fig. 1. CT image of prostate cancer

*A. CNN*

We employed specialized NN architecture designed to address the BOO problem in RT treatment planning. The network takes two inputs: anatomical data (Input 1) and a beam orientation array (Input 2). The anatomical input is composed of 2 channels representing the patient's anatomy, with one channel dedicated to the PTV and the other channel representing the weighted OARs. Notably, these channels use non-zero values based on the structure weighting obtained during the optimization step, rather than binary masks. Optional beam orientations with a resolution of 2 degrees are represented by a Boolean array of length 180, which serves as the second input. It indicates selected beam orientations in the treatment plan, while the complementary array ($BC$) denotes the remaining available beam orientations. By incorporating both arrays, the network learns more effectively when it takes into account the impact of previously chosen beam orientations.

The NN comprises three blocks of layers. The first block makes use of 3D convolutional layers, with the first two levels containing two consecutive 3-D convolutional layers (CL) with kernel sizes of $5 \times 5 \times 5$, preceding a max pooling layer of size $2 \times 2 \times 2$. The third level is made up of three consecutive 3-D CL with kernel sizes of $3 \times 3 \times 3$. The third CL's output is flattened to a 1-D layer and processed through a fully connected layer (FCL), resulting in 1024 features. The beam orientation array ($B$) and its complementary array ($BC$) are combined in the second block with the 1-D FCL. This stacked layer is subsequently subjected to an FCL, which results in 5120 features after being reformatted into 5 rows of 1024 features. The third block is made up of five stages, each beginning with two 1-D CLs with a kernel size of three and ending with an up-sampling layer. The beam information vector ($BM$) and its complement ($BMC$) are max pooled and concatenated. The final step generates a single feature array of length 180, reflecting the dual scores of beam orientations.

Each beam's dual score ($b$) reflects how adding that particular beam to the current set of selected beams ($B$) could enhance the objective function. The CNN learns from the reasoning of CG, internalizing the FMO solution without directly computing beam influence matrices. Considering the present state of the challenge, the CNN has been trained by employing an inversed and normalized dual array as the probability distribution of beam selection. The activation function used in all layers of the CNN model is the Self-Gated Rectified Linear Unit [15] for enhanced learning robustness and normalization. The Mean Squared Error (MSE) serves as the loss function during training, and the Adam optimizer with a learning rate of $1 \times 10^{-5}$ is used. To facilitate the learning process, a Supervised Learning Neural Network (SLNN) is utilized. It aims to minimize the difference between the SLNN output and the real-valued target labels provided by the teacher, in this case, the CG approach. The target labels are fitness values ($f$) which are computed from the Karush-Kuhn-Tucker (KKT) conditions [16]. The supervised training structure involves the interaction between the SLNN and the teacher (CG approach) throughout the learning process. A schematic representation of this supervised learning framework is illustrated in Figure 2.

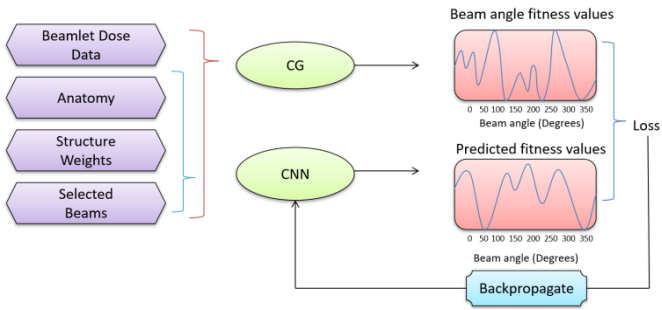

Fig. 2. Supervised training structure

To elaborate further, the optimization problem involves finding the network weights that result in the SLNN output closely matching the fitness values ($f$) obtained from the KKT conditions. The CG algorithm serves as the teacher, providing the target labels for training the SLNN. By learning from the teacher's guidance, the SLNN improves its performance and can eventually predict suitable solutions for the optimization problem. From an optimization perspective, the FMO problem, denoted as $P(B)$, can be formulated as follows:

$$min_{x,y} F(y) \qquad [1]$$

$$subject\ to\ y = \sum_{b \in B} \begin{bmatrix} D_{b,s=s_1} \\ \dots \\ D_{b,s=s_T} \end{bmatrix} x_b \qquad [2]$$

$$x_b \geq 0 \ for\ b \epsilon B \qquad [3]$$

The optimization problem involves several key components. The objective function is denoted as $F$, representing the goal to be optimized. The dose influence matrix for the $b^{th}$ beam and the $s^{th}$ structure is represented by $D_{b,s}$, which plays a crucial role in determining the dose distribution. B represents the set of selected beam orientations, and the corresponding treatment plan obtained with these selected beams is denoted as $P(B)$. In an ideal scenario without time, computation, or delivery constraints, the optimal treatment plan would utilize all possible candidate beams for treating the patient. This situation is referred to as the master problem, where B is set as Ball. The set Ball comprises all available beam orientations, denoted as $Ball = \{bj : \forall j \in J\}$.

However, practically delivering this theoretically optimal plan to the patient faces limitations. Hence, to address the optimization problem, the Column Generation (CG) method is selected. CG has proven to be effective for Beam Orientation Optimization (BOO) in RT, offering significant advantages over traditional clinical plans [17]. It is important to note that while CG does not guarantee to find the absolute optimal solution, it has demonstrated remarkable performance compared to standard clinical plans, making it a valuable and promising approach for enhancing treatment planning in RT.

*B. Column Generation*

CG is an iterative optimization approach used to approximate the solution to the limited BOO problem. The limited problem involves selecting a restricted set of beams, denoted as $B_{limit}$, until $|B_{limit}| = n$ (the desired number of beams) or until the master problem's optimality is achieved. In every iteration, CG adds one beam at a time to the set $B_{limit}$, with the primary goal of identifying the beam that provides the most significant reduction in the immediate objective value. This step-by-step process allows CG to gradually build an optimized set of selected beams, focusing on those that have the most impact in minimizing the objective function during each iteration. The objective is to efficiently find the best beams that contribute to an effective treatment plan while adhering to the given constraints. In optimization terms, CG tackles the limited problem, $P_{limit}(Ball, n)$, by solving the more comprehensive problem, $P(B)$, which encompasses the entire set of candidate beams. Through this iterative process, CG seeks to identify the most valuable beams that collectively constitute an optimal treatment plan, providing a balanced dose distribution while sparing critical structures.

*C. FMO Objective Function*

As part of the feasibility study, the FMO problem needs to be addressed for the presently chosen beam orientations [18]. For this study, we have defined $F(y)$ as a quadratic penalty function, although it's essential to note that alternative objective functions can also be considered for the FMO problem.

$$F(y) = F\left(\begin{bmatrix} y_{s=s_1} \\ \dots \\ y_{s=s_T} \end{bmatrix}\right) = \sum_{s \in S} \frac{w_s^2}{2} ||y_s - p_s||_2^2 \qquad [4]$$

In the optimization objective function, the variable $s$ represents the structure index, and $S$ denotes the set containing all structures under consideration. Each structure is assigned a user-defined structure weight, denoted as $w_s$, which represents its relative importance. Additionally, each structure is associated with a prescription dose, denoted as $p_s$, which specifies the desired radiation dose to be delivered to that particular structure. To achieve efficient optimization, we utilized the PSO and GWO algorithms, implementing them on a GPU. The PSO and GWO algorithms are well-suited for handling large-scale problems with speed and effectiveness.

*D. Optimization*

PSO and GWO are metaheuristic algorithms used to optimize the weight and bias parameters of CNNs for the FMO objective function in beam selection for RT. Let's dive into the details of each algorithm:

**Particle Swarm Optimization (PSO):** PSO, draws inspiration from the social behaviour observed in birds flocking or fish schooling [19, 20]. It models a population of particles (potential solutions) that move around the solution space. Each particle represents a set of CNN weight and bias parameters, which correspond to a potential fluence map for beam selection in RT. In PSO, particles update their positions iteratively by considering their velocity, which is influenced by two factors: their historical best position (local best) and the best position found by any particle in the entire swarm (global best) [21]. The local best represents the optimal fluence map found by a particle throughout its individual search, while the global best indicates the best fluence map discovered by any particle across the entire population. By continuously adjusting their positions based on these historical and swarm-wide bests, the particles collaborate to explore the problem space and converge towards promising solutions. Particles travel towards promising parts of the solution space based on their local and global best positions. Through this cooperative search process, PSO efficiently fine-tunes the CNN weight and bias parameters [22] to predict optimal fluence maps that minimize radiation to healthy tissues while effectively treating the tumor target.

**Grey Wolf Optimization (GWO):** GWO is inspired by the hunting behavior of grey wolves, which work together as a pack to optimize their hunting strategies [23, 24]. In GWO, three main groups of wolves are considered: alpha ($\propto$), beta ($\beta$), and delta ($\delta$). Each group represents a potential solution (fluence map) with associated CNN weight and bias parameters [25]. In the GWO, the $\propto$, $\beta$, and $\delta$ wolves represent the three best solutions found during the optimization process. The positions of these wolves are iteratively updated based on their exploration factor and a set of equations inspired by the hunting behaviour of wolves. The algorithm simulates the movement of these wolves, where their positions influence each other's movement. By leveraging the positions of the $\propto$, $\beta$, and $\delta$ wolves, the algorithm efficiently explores and refines the solution space, leading to improved optimization results over time. By mimicking the hunting process, GWO effectively explores the solution space to find fluence maps that optimize the FMO objective function for RT beam selection.

By employing PSO and GWO, RT treatment planners can efficiently explore the solution space of fluence maps, identifying optimal treatment plans that deliver the prescribed radiation dose to the tumor target while minimizing radiation exposure to healthy tissues and organs. The fine-tuned CNN weight and bias parameters lead to more precise and personalized treatment plans, resulting in improved dosimetric qualities and better treatment outcomes for cancer patients undergoing RT.

*E. Data, training, and evaluation*

In this investigation, images from 70 individuals with prostate cancer were used, each having six contours depicting the PTV, rectum, bladder, body, left, and right femoral head. The data were arbitrarily separated into two distinct sets [26]: a model training and testing sample with 57 and 13 data. The MSE was employed to evaluate the effectiveness of the CNN approach in the training phase. Following that, the trained CNN approaches forecast five beam directions for each test scenario, and the associated FMO challenges were addressed. To evaluate the performance of the CNN's predicted beam sets, the FMO solutions were compared against those generated by the CG algorithm using various metrics. This comparison aimed to assess the accuracy and efficacy of the CNN model in predicting optimal beam orientations for RT treatment planning.

- $PTV\ D_{98}$, $PTV\ D_{99}$: Represent the dose of 98% and 99% received by PTV.
- $PTV\ D_{max}$: Indicates the PTV's maximum dose.
  $PTV$ Homogeneity: Calculated as $\left(\frac{PTV\ D_2 - PTV\ D_{98}}{PTV\ D_{50}}\right)$, where $PTV\ D_2$ and $PTV\ D_{50}$ indicates the dose of 2% and 50% accepted by PTV.
- Van't Riet Conformation Number (VR): Measured as $\left(\frac{(V_{PTV} \cap V_{100\%I_{SO}})^2}{V_{PTV} * V_{100\%I_{SO}}}\right)$, where $V_{100\%I_{SO}}$ represents the volume of the isodose region which accept dose of 100%.
- R50: Calculated as $\left(\frac{V_{50\%I_{SO}}}{V_{PTV}}\right)$, where $V_{50\%I_{SO}}$ indicates the volume of the isodose region which accept dose of 50%.

By comparing these metrics between the solutions obtained by the CG algorithm and the predictions made by the CNN, the study evaluates the accuracy and effectiveness of the CNN model in optimizing beam orientations for RT. The goal is to ensure that the CNN-generated solutions lead to optimal treatment plans that provide the desired radiation dose to the target while minimizing exposure to healthy tissues, ultimately contributing to better treatment outcomes for patients with prostate cancer.

IV. RESULT AND DISCUSSION

This research emphasized the efficacy of the suggested approach by examining five-beam therapy programs for prostate cancer individuals. It is generally accepted that fewer beams in a plan make the orientation choices for those beams more crucial. For the CNN model training, we use 100 epochs. Figure 3 visually illustrates the progress of the MSE loss functions during the training process of the CNN, CNN-PSO, and CNN-GWO networks. This graph provides insights into the convergence and optimization performance of the NN models during the training phase. Figure 3 shows the performance evaluation of CNN models.

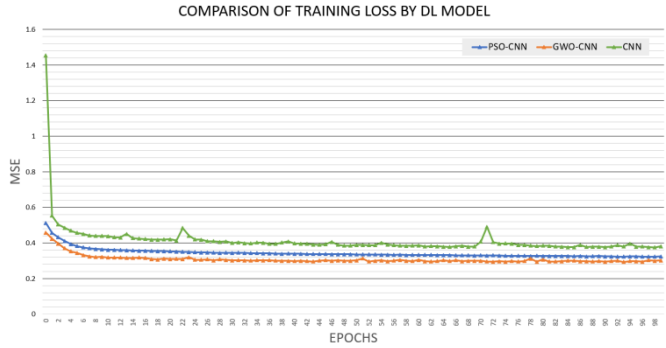

Fig. 3. CNN model performance evaluation

In Table 1, we observe the outcomes obtained by comparing the traditional CG method with the outcomes of CNN, CNN-PSO, and CNN-GWO for PTV statistics, VR, and R50 in the form of mean ± Standard Deviation (SD). The results demonstrate that all three AI-assisted methods, including CNN, CNN-PSO, and CNN-GWO, yield PTV dose statistics that are very close to those obtained by the traditional CG method. This suggests that the AI-based approaches are effective in achieving similar dose distributions within the PTV as the well-established CG method.

Furthermore, for $PTV\ D_{98}$, $PTV\ D_{99}$, and $PTV\ D_{max}$, CNN-PSO and CNN-GWO show slightly better performance with smaller SD when compared to CG. This indicates that the optimized CNN models (CNN-PSO and CNN-GWO) have the advantage of producing more consistent results, reducing uncertainty in dose coverage within the PTV compared to the traditional CNN. Regarding PTV Homogeneity, all methods perform similarly, with small differences in mean values and SD. However, both CNN-PSO and CNN-GWO demonstrate a slight improvement over CG, suggesting that the incorporation of optimization algorithms enhances the dose uniformity within the PTV. When evaluating VR and $R_{50}$, CNN-GWO emerges as the most favourable method, achieving better results compared to CG and even outperforming the optimized CNN using PSO. This indicates that the GWO shows promise in producing optimal beam orientations, leading to improved dose conformity and reduced high dose spillage, making it a preferable choice for beam selection in radiation therapy.

Table 1. Comparison of CNN, CNN-PSO, and CNN-GWO model with CG approach

| DL Model | CG | CNN | CNN-Difference | CNN-PSO | CNN-PSO-Difference | CNN-GWO | CNN-GWO-Difference |
|---|---|---|---|---|---|---|---|
| PTV $D_{98}$ | 0.9713±0.017 | 0.9671±0.014 | 0.0142±0.003 | 0.9745±0.012 | -0.0032±0.005 | 0.9713±0.021 | 0±0.004 |
| PTV $D_{99}$ | 0.9513±0.004 | 0.9587±0.142 | -0.0174±0.138 | 0.9518±0.041 | -0.0005±0.037 | 0.9511±0.013 | 0.0002±0.009 |
| PTV $D_{MAX}$ | 0.8647±0.021 | 0.8410±0.087 | 0.0237±0.066 | 0.8612±0.037 | 0.0035±0.016 | 0.8649±0.084 | -0.0002±0.063 |
| PTV Homogeneity | 0.062±0.037 | 0.078±0.135 | -0.016±0.098 | 0.060±0.031 | -0.008±0.006 | 0.062±0.0045 | 0±0.0325 |
| VR | 0.835±0.0074 | 0.8014±0.0165 | 0.0336±0.0091 | 0.875±0.098 | -0.04±0.0906 | 0.831±0.0071 | 0.004±0.0003 |
| $R_{50}$ | 4.178±0.745 | 4.412±0.613 | -0.234±0.132 | 4.124±0.621 | 0.054±0.124 | 4.172±0.685 | 0.006±0.06 |

In Figure 4, we present visual illustrations of the final FMO solutions. These solutions were obtained by selecting beam orientations using three distinct methods: CG, CNN, and the optimized CNN using GWO and PSO. The beam selection and resulting FMO solutions are influenced by the weights assigned to the anatomical structures during the optimization process. The Dose-Volume Histogram (DVH) graphs in Figure 4.a present the treatment plans' dose distribution for the PTV, rectum, and bladder. Figure 4.b illustrates the DVH curves for the left femur, right femur, and the patient's body as a whole. Remarkably, the DVH curves for the anatomical structures obtained using CG, and CNN-GWO methods exhibit significant similarity. This similarity indicates that the radiation treatment plans generated by CG and CNN-GWO have a comparable impact on the patient's body. This consistency in DVH curves was expected from the CNN solution. The similarity in DVH curves suggests that both the CG method and the optimized CNN-GWO approach achieve comparable dose distributions for the organs and structures being considered. This finding further validates the effectiveness of the CNN-GWO model as an alternative to the traditional CG method in achieving similar and clinically acceptable treatment plans.

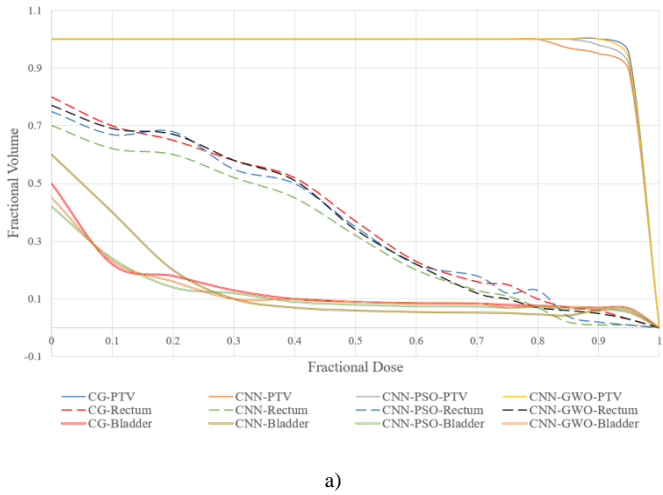

a)

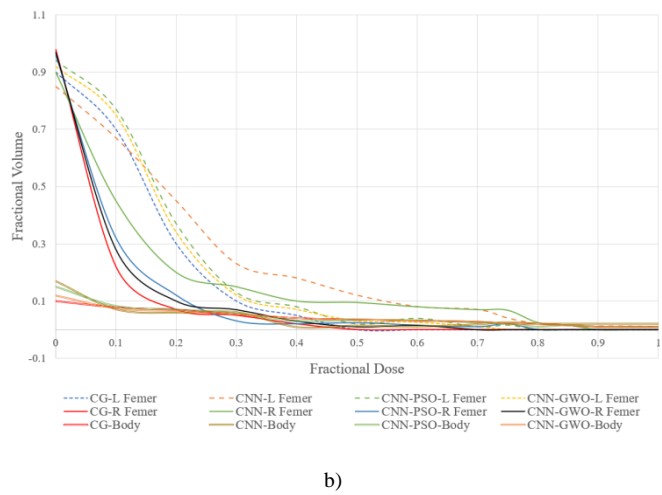

b)

Fig. 4. DVH graph

## V. CONCLUSION

In this study, we have successfully designed and demonstrated optimization strategies for beam direction and dose distribution selection in radiotherapy planning using advanced AI techniques. Our approach is superior because we use a CNN to selectively calculate dosage effect matrices for only the beams that have been chosen for treatment. This significantly reduces planning time while guaranteeing that treatment plan quality is on par with conventional CG approaches. The trained CNN demonstrates its capability by making informed decisions on beam orientations solely based on patients' anatomical information, such as contoured images and structure weights. In contrast, the CG method demands extensive dose influence matrix calculations for all potential beam orientations, which is computationally intensive and time-consuming. The CNN, CNN-PSO, and CNN-GWO models predict sets of five beam orientations in just 1.5, 2.1, and 2.2 seconds, respectively. Subsequently, the FMO process takes only a few minutes. In total, the CNN, CNN-PSO, and CNN-GWO methods require 5 minutes and 20 seconds, 6 six minutes and 12 seconds, and 6 minutes and 47 seconds, respectively, to generate the optimal treatment plan. To determine the dose for every possible beam, the CG approach takes more than three hours. The successful integration of our AI-driven optimization strategies has the potential to revolutionize radiotherapy planning, enhancing treatment efficiency, and patient care. The considerable time saved by the CNN-based models makes them highly suitable for clinical implementation, enabling radiation oncologists to deliver precise and tailored radiotherapy treatment plans with reduced waiting times for patients.

While our optimization strategies have shown promising results, there are several avenues for future research and development in this domain. Firstly, extending the application of our approach to other cancer types and anatomical sites would broaden its scope and applicability. Investigating the model's performance in complex clinical scenarios and accounting for additional treatment constraints will offer valuable insights into its adaptability and versatility. Integrating other AI techniques, such as reinforcement learning or generative adversarial networks, could further enhance the optimization process and generate even more sophisticated treatment plans. Combining multiple AI models or hybrid strategies may lead to novel approaches for optimizing beam direction and dose distribution in radiotherapy planning.